\documentclass[aps,twocolumn,floats,floatfix,showpacs]{revtex4}

\usepackage{graphicx,amsmath,amssymb,multirow,rotate,color}

\usepackage{ae,aecompl}
\usepackage[latin1]{inputenc}

\begin{document}

\title{Accumulation horizons and period-adding in 
       optically injected semiconductor lasers}

\author{Cristian \surname{Bonatto}}
\affiliation{Instituto de F\'\i sica, Universidade Federal do 
Rio Grande do Sul, 91501-970 Porto Alegre, Brazil}
\author{Jason A.C.~\surname{Gallas}}
\affiliation{Instituto de F\'\i sica, 
             Universidade Federal do Rio Grande do Sul, 
             91501-970 Porto Alegre, Brazil}
\date{\today}
\begin{abstract}
We study the hierarchical structuring of islands of stable
periodic oscillations inside chaotic regions in phase diagrams of 
single-mode semiconductor lasers with  optical injection.
Pha\-se dia\-grams display remarkable {\it accumulation horizons}: 
boundaries formed by the accumulation of infinite cascades
of self-similar islands of periodic solutions of 
ever-increasing period.
Each cascade follows a specific pe\-ri\-od-ad\-ding route.
The riddling of chaotic laser phases by such networks of periodic
solutions may compromise applications operating with chaotic signals 
such as e.g.~secure communications.
\end{abstract}



\maketitle

\begin{figure*}[bth]
\includegraphics[width=5.7cm]{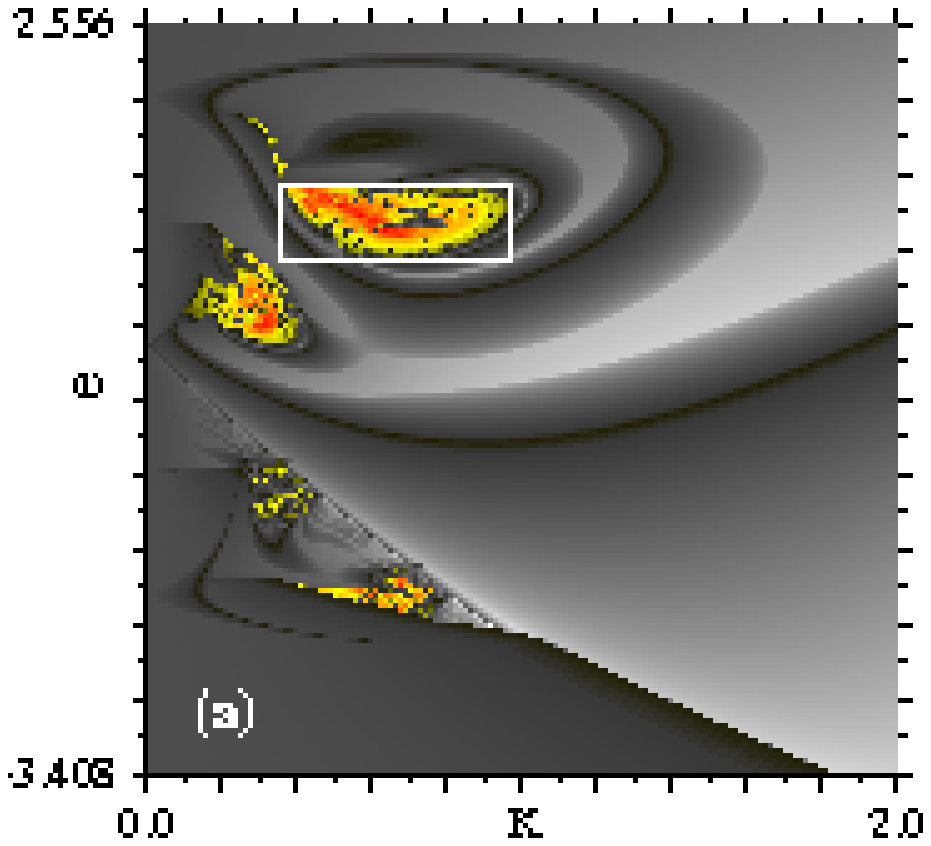}
\includegraphics[width=5.7cm]{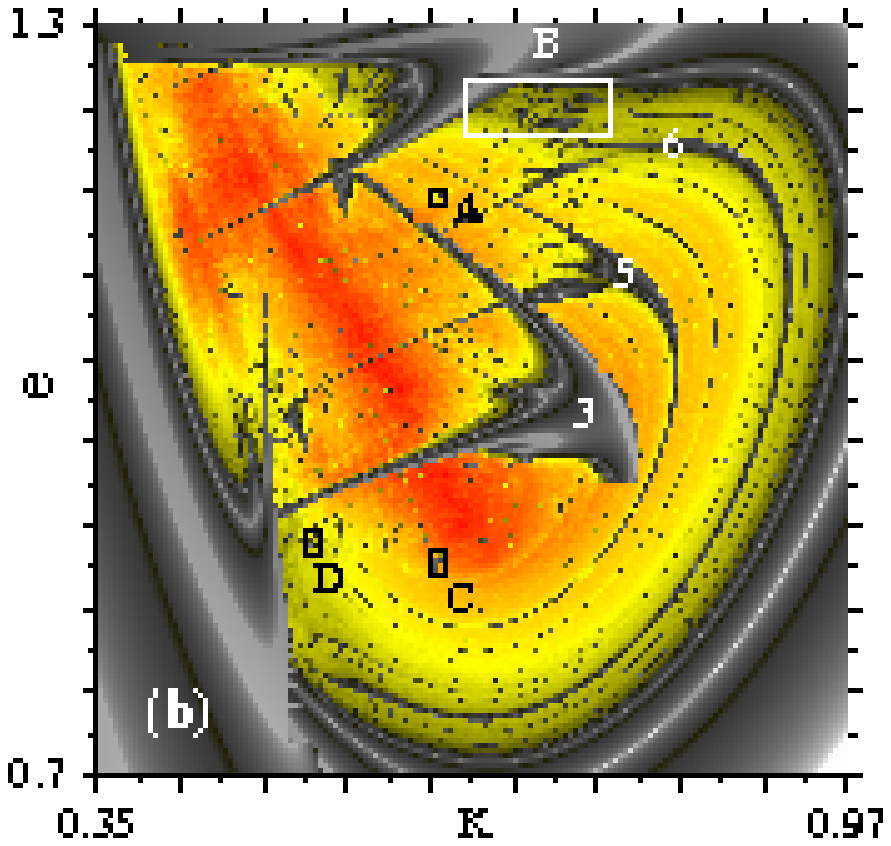}
\includegraphics[width=5.7cm]{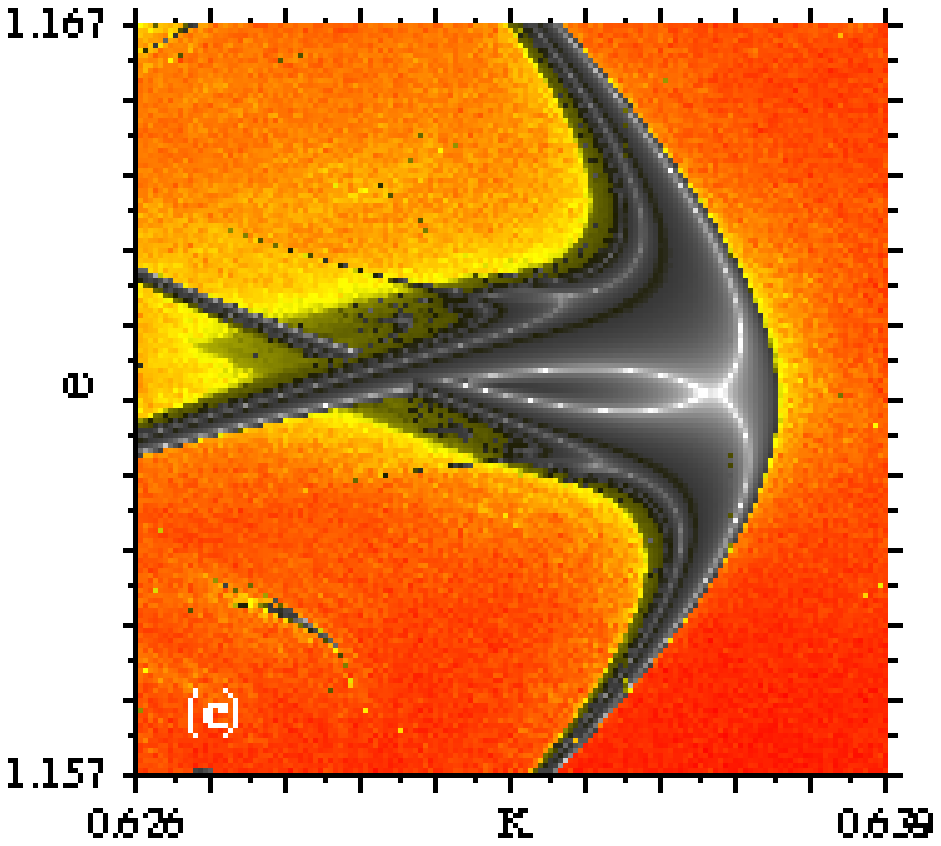}
\caption{\protect (Color online)  Phase diagrams quantifying both
regularity (darker shadings) and chaos (colors; lighter shadings).
(a) Global view.
(b) Magnification of box in (a), for positive detuning:
    Numbers denote quantity of peaks in a period of the laser intensity.
    Boxes A, B, C, and D  are shown magnified in the next figures.
(c) Magnification of the period-9 island inside 
    box A in (b) showing a structure also found in CO$_2$ lasers \cite{bgg05}.
    Red denotes ``stronger'' chaos (more positive Lyapunov exponents).
    NOTE: the original high-resolution of all figures was greatly reduced to
    comply with arXiv limitations.
}
\label{fig:fig01}
\end{figure*}

\begin{figure*}[bth]
\includegraphics[width=5.7cm]{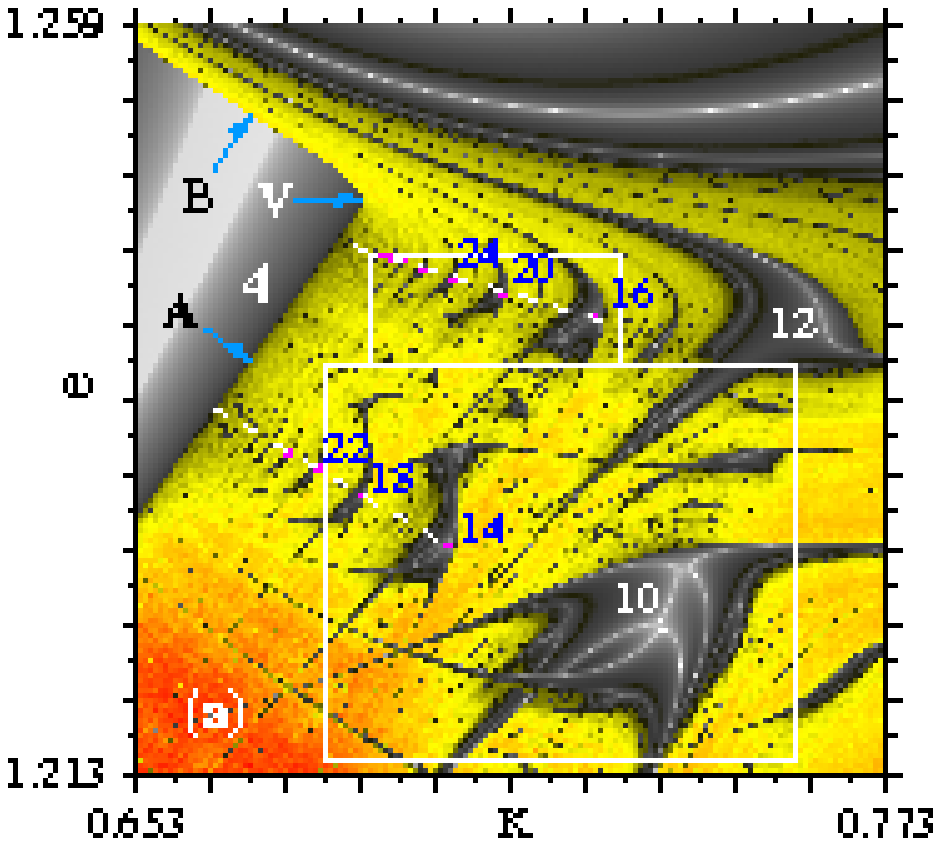}
\includegraphics[width=5.7cm]{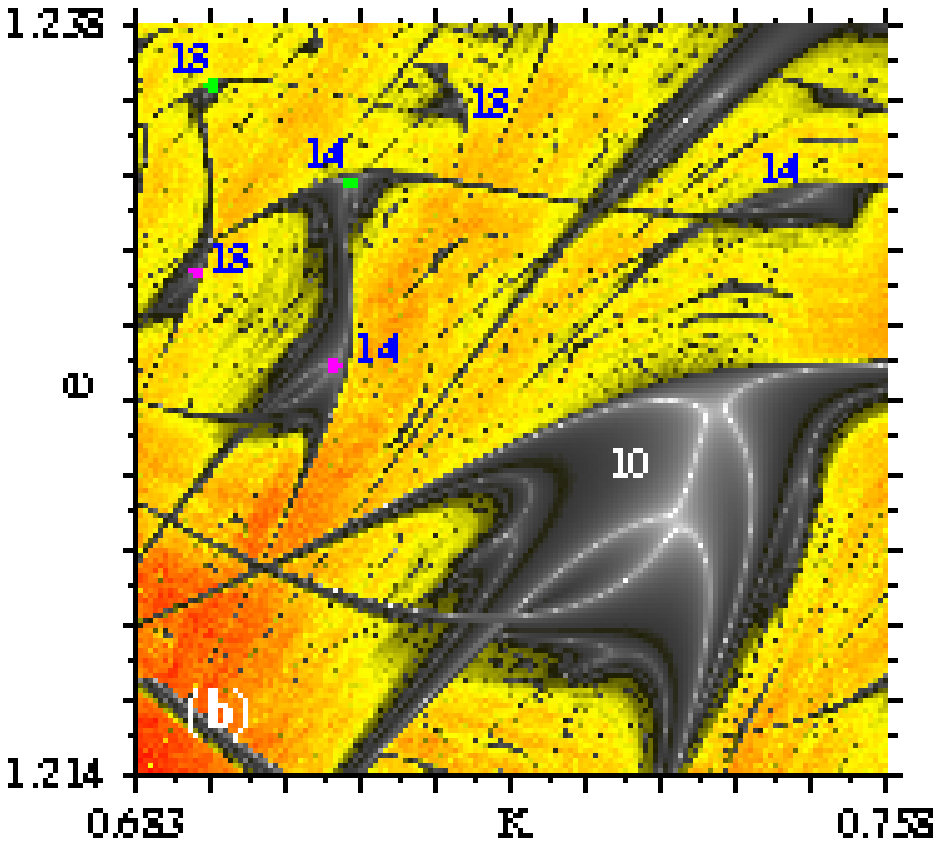}
\includegraphics[width=5.7cm]{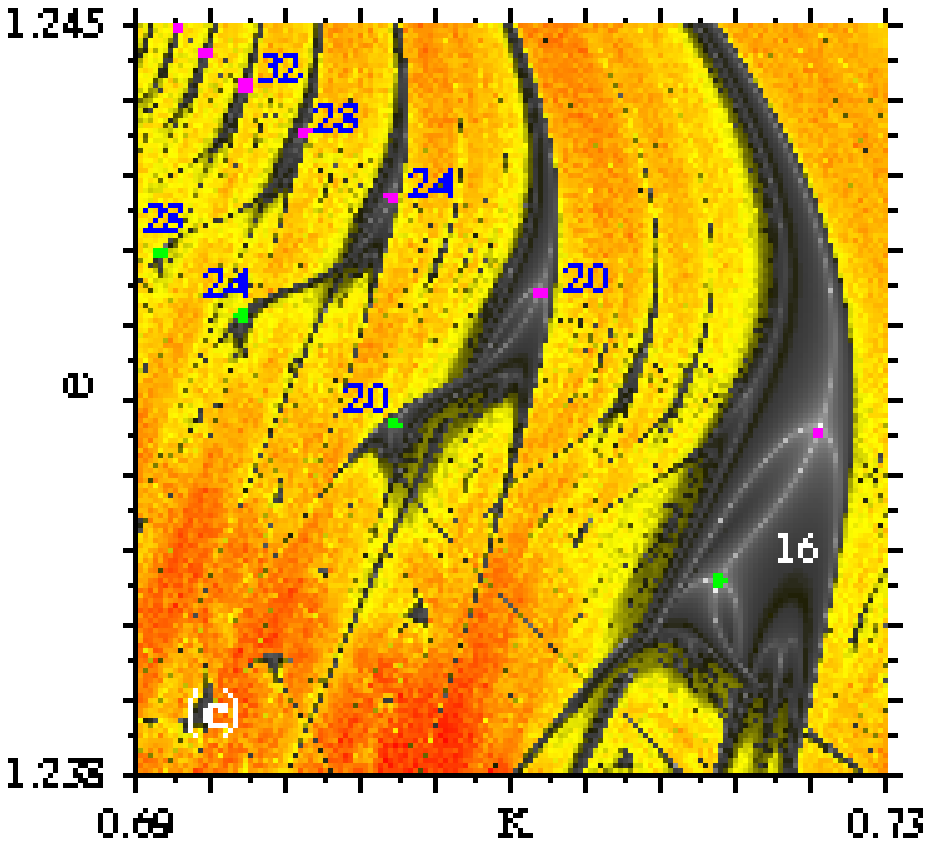}
\vspace{0.2cm}
\caption{\protect (Color online)  Accumulations of islands
of periodic solutions (darker shadings) 
embedded inside a chaotic phase (yellow-red lighter shadings).
(a) Series of islands converging toward a line segment, marked A, 
    forming an  {\it accumulation horizon}.
    ``Legs of periodicity'' accumulate parallel to line B.
    Curves A and B meet at vertex V.
    Bifurcation diagrams along dotted lines,
    shown in Fig.~\ref{fig:fig03}, display 
    period-adding cascades converging to horizon A,
    the boundary of a four-peak domain, as indicated.
(b) Genesis and separation of  distinct
    $10\to14\to18\to\dots$ period-adding cascades.
(c) Similar genesis and separation as in (b) but for 
    distinct $(12)\to16\to20\to24\to\dots$ cascades.
Numbers refer to the quantity of peaks in the laser intensity.
}
\label{fig:fig02}
\end{figure*}

Semiconductor lasers are key components for progress in 
many areas \cite{book1,book2}, e.g.~in the investigations of signatures 
of wideband chaos  synchronization \cite{book1,shore06}, 
for secure communication \cite{book2,ohtsubo02},
and for ultrahigh-density optical memories \cite{leuchs03}.
Such compact lasers represent 99.8\% of the world market for lasers
in terms of units sold per year \cite{lf04}, a market of 
600 million units sold in 2003/04. 
An important quality of semiconductor lasers is their rich 
nonlinear response when subjected either to optical injection, to 
optical feedback, or to modulations. 
Lasers with optical injection have attracted
much attention in recent years, experimentally as well as from
 theoretical and numerical points of view.
The complex  phenomenology and intricate structure of bifurcations 
as a function of the injected intensity and frequency detuning
were summarized in a recent survey \cite{report}.

The present literature indicates a good overall agreement 
between theory and experiments  \cite{wie2002}.
For instance,  calculations and numerical simulations
predict  intricate laser behaviors, including stable
periodic oscillations inside regions characterized by 
chaotic laser signals  \cite{gao1999,hwa2000}.
More recently,
Fordell and Lindberg \cite{for2004} and  Chlouverakis and Adams 
\cite{chl2003} reported  diagrams obtained by 
numerical integration of the rate equations for an optically
injected semiconductor laser showing some
islands of periodic laser signals  embedded in a sea of chaos.
These important findings raise an interesting question concerning 
the precise structuring of laser  chaotic phases.
In fact, this question is the tip of a much wider problem
that we consider here.

Phase diagrams for discrete-time models described by mappings 
are common nowadays \cite{shrimps93,hunt}.
But the much more difficult problem of building detailed
phase diagrams for models ruled by sets of nonlinear 
{\it differential equations\/} has been much less investigated.
Of course, diagrams recording complex bifurcations 
and providing valuable insight for a few of the lowest periods 
have been obtained in a number of in-depth bifurcation
studies using  powerful continuation methods \cite{report,ref1,ref2,ref3}.
However, complete diagrams, discriminating simultaneously
regions of arbitrarily high periods  and regions with chaotic phases, remain
essentially unexplored for continuous-time autonomous models.
This is the problem we attack here.

Our numerical simulations revealed surprising regularities existing
{\it inside the chaotic phases\/} of the laser.
As illustrated in  Figs.~\ref{fig:fig01} and \ref{fig:fig02},
the parameter space has wide regions characterized by chaotic solutions.
These chaotic {\it phases\/}  contain both  single accumulations 
as well as accumulations of accumulations.
More specifically,  
chaotic laser phases are riddled  with infinite sequences of 
period-adding cascades,  
each one converging toward  curves that look  simple (structureless),
denoted  ``accumulation horizons'', for simplicity.
One example is indicated by the arrow marked  A in Fig.~\ref{fig:fig02}a. 
From a theoretical point of view,
a key novelty here is that the differential equations ruling the
laser are {\it autonomous equations}, i.e., they do not involve time
explicitly. 
Thus, the remarkable organization of the parameter space reported here
must originate from
an {\it intrinsic interplay between variables and parameters\/}
of the laser.
We found
accumulation horizons  to exist abundantly also in electronic circuits,  
atmospheric and  chemical oscillators, and  several
other  systems \cite{novo}.
To fix ideas, here we focus just on the laser case.
Incidentally, we mention that accumulation cascades in 
semiconductor lasers have been investigated  by 
Krauskopf and Wieczorek \cite{ref3} quite recently.
However, their accumulations are of a very different 
nature than ours \cite{difer}.

The laser we consider is a single-mode semiconductor laser
subjected to monochromatic optical injection,  governed by the
standard rate equations for the complex laser field
$E=E_x+iE_y$ 
and a population inversion $n$ rescaled such that \cite{wie2002}
\begin{subequations}
\label{model}
\begin{eqnarray}
\dot{E} &=& K + \Big(\tfrac{1}{2}(1+i\alpha)n-i\omega\Big)E,\label{model_a}\\
\dot{n} &=& -2\Gamma n - (1+2 B n)(|E|^2-1).   \label{model_b}
\end{eqnarray}
\end{subequations}
Here, the interesting control parameters are $K$, the intensity of 
the injected field, and $\omega$, the detuning frequency.
As usual \cite{wie2002}, we fix
$B=0.0295$, $\Gamma=0.0973$, and $\alpha=2.6$.

Figure \ref{fig:fig01} illustrates typical high-resolution phase dia\-grams 
obtained by computing the spectra of Lyapunov exponents 
on a $900\times900$ grid of equally spaced parameters
for Eqs.~(\ref{model_a}) and (\ref{model_b}), 
integrated with a standard
fourth-order Runge-Kutta scheme with a fixed step size $h=0.01$.
Each grid point color-codifies
the magnitude of the largest nonzero exponent:
negative exponents (indicating periodic solutions) were colored with
gray shadings (black indicates zero, white the most negative values),
while positive exponents (marking chaotic laser signals) 
are indicated in a continuously changing yellow-red scale
(lighter shadings in black and white printers; check online figures).
The color scale of individual phase diagrams was 
renormalized to span each diagram.
Figure \ref{fig:fig01}a displays the same parameter region investigated
recently by Wieczorek  {\it et al } \cite{wie2002}.  
To convert  $\omega$ into Ghz, multiply it by $4.6948$.
Our figure corroborates the low-period bifurcation 
boundaries reported recently \cite{wie2002}  and, more importantly,  shows
additional details and regularities not observed before, like
e.g., the inner structuring of periodicity domains, the regions where
recurring self-similar organizations occur and where they fail to exist.
Our figures reveal details which are very hard
(if not impossible) to come by using continuation methods.

Islands of regular laser oscillations  in semiconductor lasers were 
measured by  Eriksson and Lindberg in recent experiments \cite{eri01,eri02}.
First, they were able to identify a period-3 island
by tuning the injection intensity for three fixed values of the
frequency detuning \cite{eri01}.
Then, by repeating measurements for finer detuning intervals,
they cleverly managed to characterize a few islands of 
low period \cite{eri02}.
Figure \ref{fig:fig01}b  corroborates such low-periodic islands and
shows a myriad of additional islands of ever-increasing periods
as discussed below. 
It also displays several novel features,
in particular the existence of self-similarities of various kinds.
Figure \ref{fig:fig01}c displays an island with the familiar 
shrimp-shape \cite{shrimps93} 
recorded when varying two parameters simultaneously 
(codimension-two phenomenon).
Although well-known in {\it discrete}-time dynamical systems, 
this peculiar shape was observed only recently 
in a non-autonomous  {\it continuous}-time system, namely in 
CO$_2$ lasers \cite{bgg05}.

A series of unexpected and striking  accumulation networks 
may be easily recognized from Fig.~\ref{fig:fig02}, 
presenting successively magnified views of box B 
in Fig.~\ref{fig:fig01}b.
Embedded in the chaotic region there are regular and abundant networks 
of stable islands of periodic laser signals with unbounded periodicities.
As Fig.~\ref{fig:fig02}a  shows,
the parameter networks living in the chaotic region 
bridge  periodic laser signals of increasingly higher periodicities,
which converge systematically 
toward well-defined and characteristic  accumulation boundaries
or horizons.
As indicated schematically by the numbers in Fig.~\ref{fig:fig02}a,
when moving along the dark central bodies of the islands one
observes series of {\it period-adding cascades\/} of bifurcations,
a characteristic signature of the experimentally  elusive and rather 
challenging homoclinic route to 
chaos \cite{swi83,hom1,hom2,shil92,hc93,rm02,zeb03,ic05}.
Note that the periodicity in these cascades increases by $4$, the
same periodicity characterizing the region of periodicity that exists
to the left of the accumulation boundary.

That periodicity organizes along specific directions in parameter space 
is a well-known fact for {\it discrete}-time dynamical 
systems \cite{shrimps93}. 
But that this is also  true for {\it continuous}-time dynamical 
system is  made obvious now by Fig.~\ref{fig:fig02}.
A feature not yet reported for  discrete-time systems 
is the original way in which individual
period-adding bifurcation cascades are born \cite{novo}.
As shown by Fig.~\ref{fig:fig02}b, the single period-10 
structure (containing the pair of quasi-osculating white spines)
{\it splits\/} into two essentially separated 
shrimplike  structures \cite{shrimps93} 
as the period increases. This mechanism
leads to separate cascades that quickly give the impression
of being totally uncorrelated because of the very strong compression
experienced by the islands as the period increases more and more
without bound. Here, white spines mark loci of the most negative
Lyapunov exponents, being  loosely equivalent to the
superstable loci familiar from  discrete-time dynamical systems.
The splitting process involves several specific metric properties,
for instance, the parameter separation of the islands
accumulates to specific values while their volume decreases
regularly with characteristic exponents.

\begin{figure}[thb]
\includegraphics[width=8.1cm]{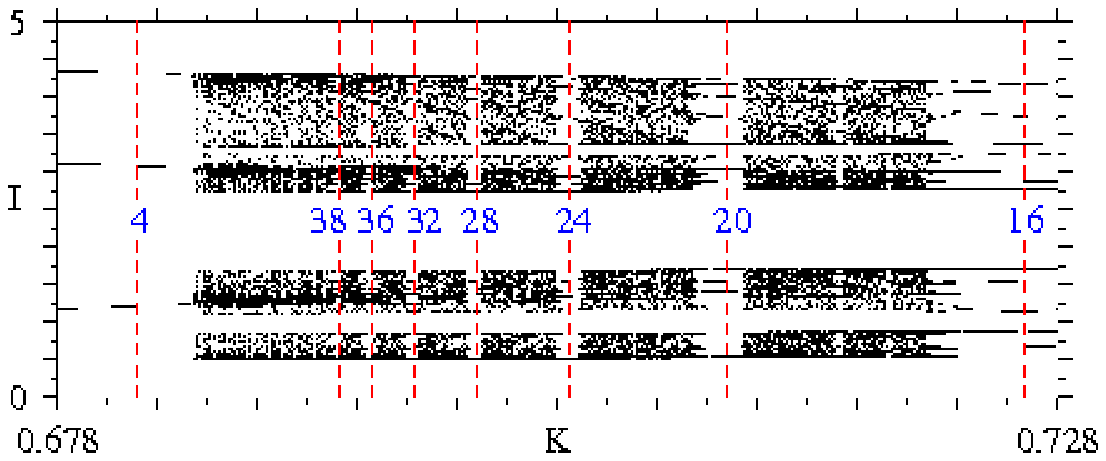}\\
\includegraphics[width=8.1cm]{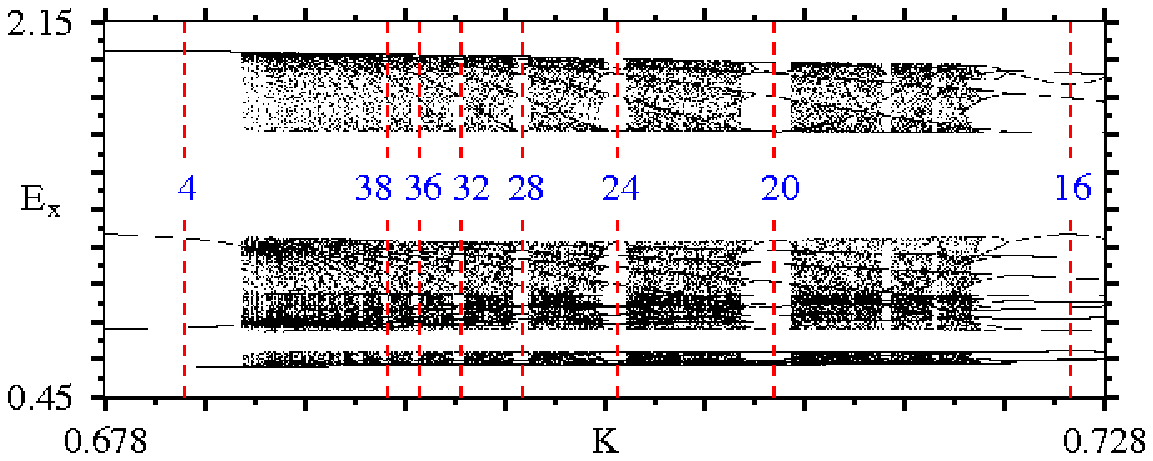}\\
\includegraphics[width=8.1cm]{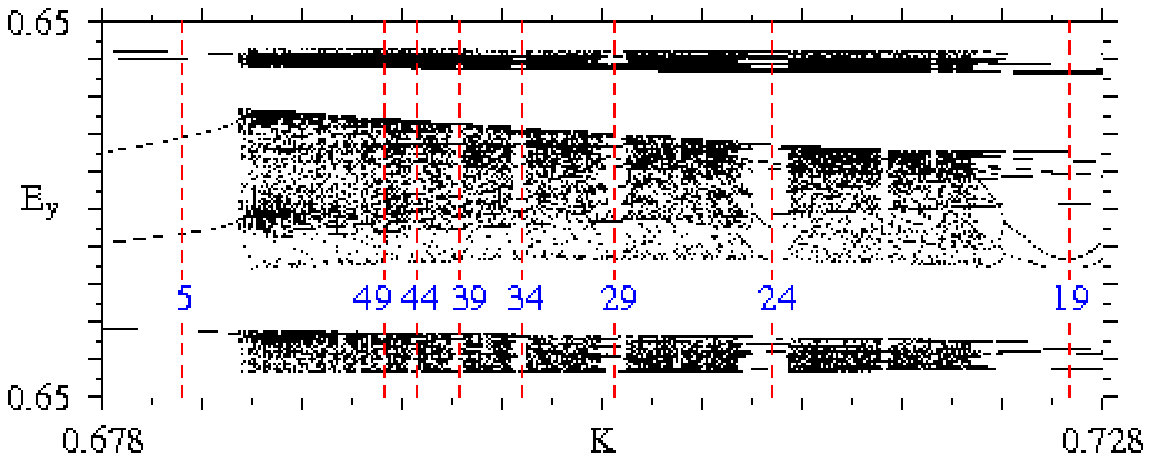}\\
\includegraphics[width=8.1cm]{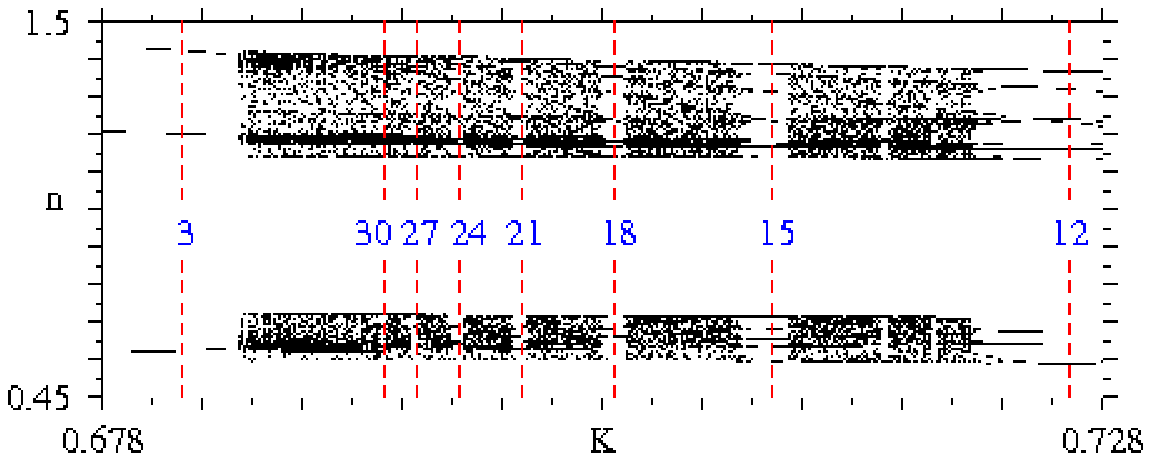}
\caption{\protect (Color online) Bifurcation diagrams showing that 
the number of peaks of the signals depends on the physical quantity
being considered. The number of peaks of  laser intensity 
$I \equiv {E_x^2 + E_y^2}$ coincides with that in $E_x$.}
\label{fig:fig03}
\end{figure}

\begin{figure}[htb]
\includegraphics[width=4.25cm]{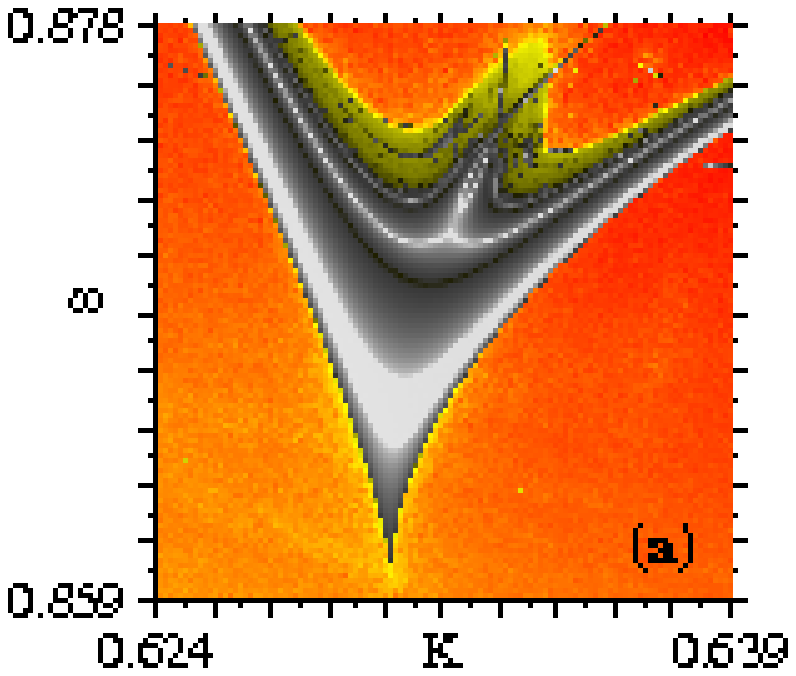}
\includegraphics[width=4.25cm]{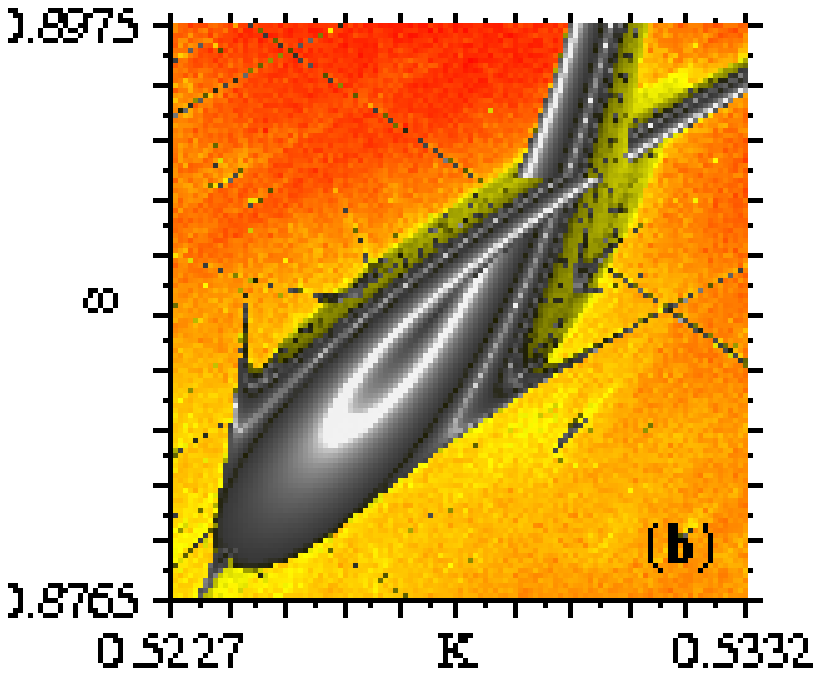}
\caption{\protect (Color online) Magnification of boxes C and D
 in Fig.~\ref{fig:fig01}b, showing typical islands or 
 stable periodic orbits with the same shapes found recently 
 in a completely different and novel scenario:  
 systems {\it without critical points} (see text).
Color coding as in Figs.~\ref{fig:fig01} and \ref{fig:fig02}.}
\label{fig:fig04}
\end{figure}

The bifurcation diagrams in Fig.~\ref{fig:fig03},
obtained when moving along the upper dotted path 
in Fig.~\ref{fig:fig02}a,   show  period-adding cascades
with the  characteristic alternation of chaos  
and  periodicity \cite{swi83,hom1,hom2,shil92,hc93,rm02,zeb03,ic05}.
Numbers labeling periodic windows refer to the number of 
peaks present in one  period of the respective variable.
Note the striking  fact that
{\it different variables display different number of peaks}. 
Since the number of peaks is usually taken to label the
``period'' of oscillation, one sees that such labels are not unique
but {\it depend on the variable used to count the peaks}.
Note that, independently of the variable selected, the number of
periods increases by an amount equal to the number of peaks 
characterizing the leftmost window, toward which
the period-adding cascades accumulate.

As a last noteworthy  result found in 
semiconductor lasers, Fig.~\ref{fig:fig04}  illustrates
islands of regular signals having the same  exquisite {\it shapes\/} 
found  very recently in a rather different scenario: 
in a discrete-time dynamical system with no critical points, 
i.e., in a system not obeying the 
Cauchy-Riemann conditions \cite{eg06,egPLA}.
Such striking shapes exist abundantly
in the lower portion of Fig.~\ref{fig:fig01}b.
Thus, semiconductor lasers open the way to investigate experimentally
novel and sophisticated mathematical behaviors arising from
holomorphic dynamics not ruled by {\it critical points}, 
so far believed to be the key
players in the dynamics of complex functions \cite{eg06}.

In summary, chaotic phases of optically 
injected semiconductor lasers contain  peculiar accumulation 
boundaries and networks formed by stable periodic solutions.
Since extended domains of ``clean'' chaos are critical for a number 
of laser applications \cite{book1,book2}, 
these regularities  need to be duly taken 
into account in applications that depend on the existence of wide
regions of smooth and continuous chaos,
such as secure communication with chaos.
Although  we concentrated on the case $\alpha=2.6$,
representative of the relatively low values more frequently addressed 
in the literature, 
larger islands exist for higher values of $\alpha$, say $\alpha\simeq 6$,
making them easier to observe experimentally.
Accumulation horizons exist also in 
other laser systems, e.g., in CO$_2$ lasers with feedback, and in  other
sets of differential equations \cite{novo}.

The accumulation networks reported here pose an interesting question:
In sharp contrast with discrete dynamical system, 
where periodicity varies discretely (``quantized''),
an appealing new possibility afforded by lasers is to
study how periodicities defined by continuous real numbers
evolve and organize in phase diagrams when several parameters 
are tuned simultaneously.
Such investigations should not be too difficult to perform
numerically.
As a last remark, we briefly mention that the  
{\it alternating period-chaotic\/} 
sequences familiar from period-adding 
cascades \cite{swi83,hom1,hom2,shil92,hc93,rm02,zeb03,ic05}
are in fact an illusory artifact of considering too restricted slices
cutting  very regular structures in parameter space  \cite{novo}.
The proper unfolding of this phenomenon requires tuning at least
two parameters, i.e., is a phenomenon visible only in
codimension-two or higher.

The authors thank  CNPq,
Brazil, for a Doctoral Fellowship 
(CB) and a Senior Research Fellowship (JACG).
This work was also supported by the AFOSR, 
Grant FA9550-07-1-0102.

\end{document}